\documentclass{article}
\usepackage{ijcai16}
\usepackage{graphicx}
\usepackage{psfrag}
\usepackage{subfigure}
\usepackage{amsmath,amssymb}
\usepackage{array}
\usepackage{times}
\usepackage{xfrac}
\usepackage{url}

\title{
Protein Secondary Structure Prediction Using Cascaded Convolutional and\\ Recurrent Neural Networks}
\author{
Zhen Li\hspace{40mm} Yizhou Yu\\
Department of Computer Science, The University of Hong Kong\\
zli@cs.hku.hk, yizhouy@acm.org}
\begin{document}

\maketitle

\begin{abstract}
Protein secondary structure prediction is an important problem in bioinformatics. Inspired by the recent successes of deep neural networks, in this paper, we propose an end-to-end deep network that predicts protein secondary structures from integrated local and global contextual features. Our deep architecture leverages convolutional neural networks with different kernel sizes to extract multiscale local contextual features. In addition, considering long-range dependencies existing in amino acid sequences, we set up a bidirectional neural network consisting of gated recurrent unit to capture global contextual features. Furthermore, multi-task learning is utilized to predict secondary structure labels and amino-acid solvent accessibility simultaneously. Our proposed deep network demonstrates its effectiveness by achieving state-of-the-art performance, i.e., $\bf 69.7\%$ Q$8$ accuracy on the public benchmark CB$513$, $\bf 76.9\%$ Q$8$ accuracy on CASP$10$ and $\bf 73.1\%$ Q$8$ accuracy on CASP$11$. Our model and results are publicly available\footnote{\url{https://github.com/icemansina/IJCAI2016}}.
\end{abstract}

\section{Introduction}

Accurately and reliably predicting structures, especially $3$D structures, from protein sequences is one of the most challenging tasks in computational biology, and has been of great interest in bioinformatics~\cite{yaseen2014}. Structural understanding is not only critical for protein analysis, but also meaningful for practical applications including drug design \cite{Noble2004}.
Understanding protein secondary structure is a vital intermediate step for protein structure prediction as the secondary structure of a protein reflects the types of local structures (such as $3_{10}-$helix and $\beta-$bridge) present in the protein. Thus, an accurate secondary structure prediction significantly reduces the degree of freedom in the tertiary structure, and can give rise to a more precise and high resolution protein structure prediction~\cite{yaseen2014,zhoujian2014,wang2011}.

The study of protein secondary structure prediction dates back to $1970$s. In the $1970$s, statistical models were frequently used to analyze the probability of specific amino acids appearing in different secondary structure elements~\cite{chou1974}. The Q$3$ accuracy, i.e., the accuracy of three-category classification: helix (H), strand (E) and coil (C), of these models was lower than $60\%$ due to inadequate features. In the $1990$s, significant improvements were achieved by exploiting the evolutionary information of proteins from the same structural family~\cite{rost1993} and position-specific scoring matrices~\cite{Jones1999}. During this period, the Q$3$ accuracy exceeded $70\%$ by taking advantage of these features. However, progress stalled when it came to the more challenging $8$-category classification problem, which needs to distinguish among the following $8$ categories of secondary structure elements: $3_{10}-$helix (G), $\alpha-$helix (H), $\pi-$helix (I), $\beta-$strand (E), $\beta-$bridge (B), $\beta-$turn (T), bend (S) and loop or irregular (L)~\cite{zhoujian2014,yaseen2014template}. In the $21$-st century, various machine learning methods, especially artificial neural networks, have been utilized to improve the performance, e.g., SVMs~\cite{Hua2001}, recurrent neural networks (RNNs)~\cite{pollastri2002}, probabilistic graphical models such as conditional neural fields combining CRFs with neural networks~\cite{wang2011}, generative stochastic networks~\cite{zhoujian2014}.

\begin{figure}[t]
{\includegraphics[width=1\columnwidth]{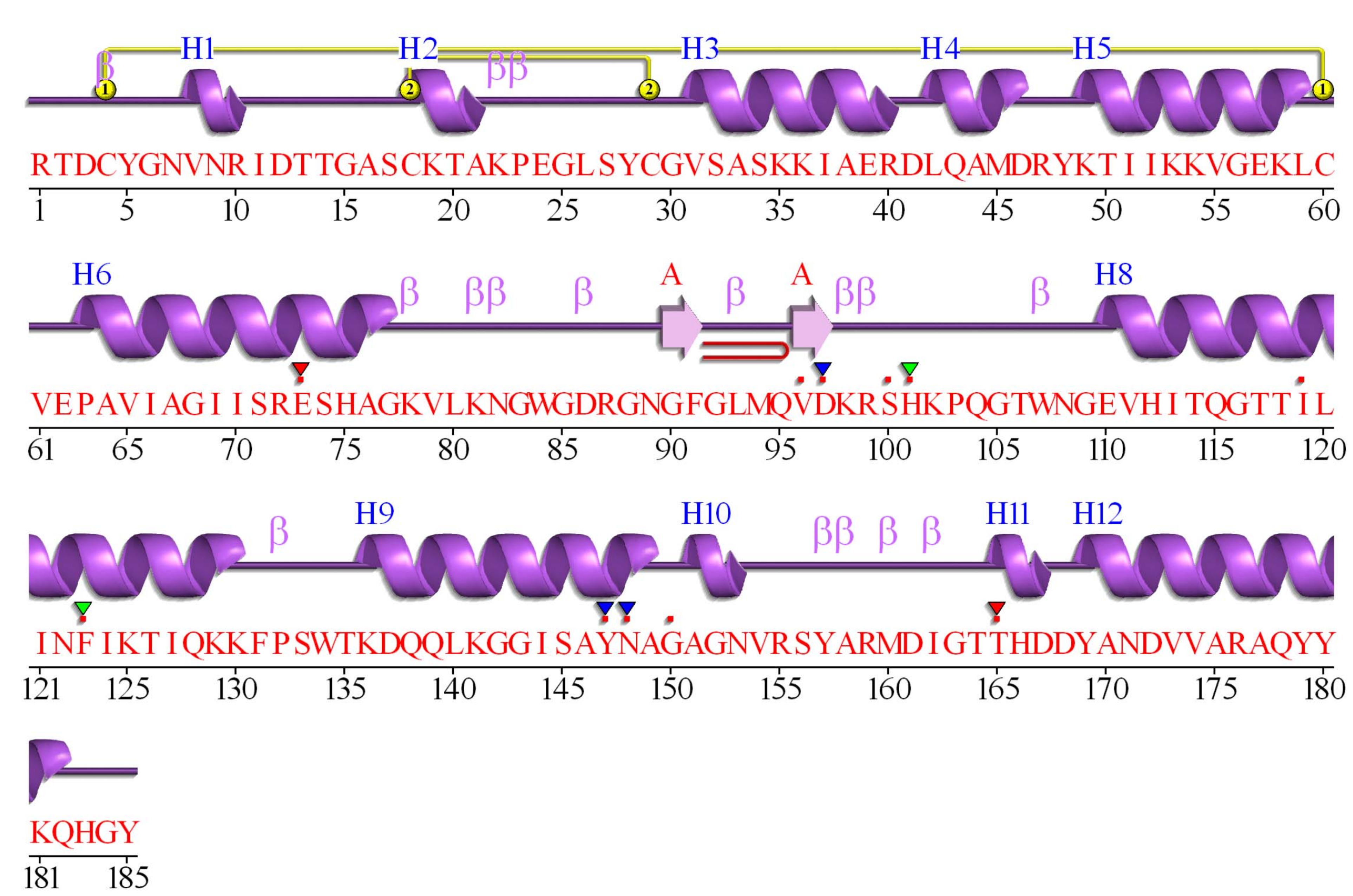}}
\caption{The amino acid sequence and its corresponding $3$-state secondary structure of PDB $154$L with UniProtKB accession number (P$00718$), which consists of $185$ residues.}\label{fig1.secondary}%
\end{figure}

It is well known that local contexts are critical for protein secondary structure prediction. Specifically, the secondary structure category information of the neighbours of an amino acid are the most effective features for classifying the secondary structure this amino acid belongs to. For instance, in Fig.\ref{fig1.secondary}, the $18$th to $21$th amino acids in PDB $154$L~\cite{simpson1983complete} (obtained from the publicly available protein data bank~\footnote{\url{http://www.rcsb.org/pdb/explore/explore.do?structureId=154L}}) are likely to be assigned the same secondary structure label given their neighbours' information.
Convolutional neural networks (CNN)~\cite{lecun1998gradient}, a specific type of deep neural networks using translation-invariant convolutional kernels, can be applied to extracting local contextual features and have proven to be effective for many natural language processing (NLP) tasks~\cite{yih2011learning,zhang2015character}. Inspired by their success in text classification, in this paper, CNNs with various kernel sizes are used to extract multiscale local contexts from a protein sequence.

On the other hand, long-range interdependency among different types of amino acids also holds vital evidences for the category of a secondary structure, e.g., a $\beta-$strand is steadied by hydrogen bonds formed with other $\beta-$strands at a distance~\cite{zhoujian2014}. For instance, also in Fig.\ref{fig1.secondary}, the $4$th and $60$th amino acids can be determined to share the same secondary structure label given the disulphide bond annotated as link $1$.
Similar to CNNs, recurrent neural networks (RNNs) are another specific type of neural networks with loop connections. They are designed to capture dependencies across a distance larger than the extent of local contexts.
In previous work~\cite{Babaei2010}, RNN models could not perform well on protein secondary structure prediction partially due to the difficulty to train such models. Fortunately, RNNs with gate and memory structures, including long short term memory (LSTM)~\cite{Schmidhuber1997}, gate recurrent units (GRUs)~\cite{cholearning}, and JZ3 structure~\cite{ilya2015}, can artificially learn to remember and forget information by using specific gates to control the information flow.
In this paper, we exploit bidirectional gate recurrent units (BGRUs) to capture long-range dependencies among amino acids from the same protein sequence.


In summary, the main contributions of this paper are as follows.
 \begin{itemize}
\item We propose a novel deep convolutional and recurrent neural network (DCRNN) for protein secondary structure prediction. This deep network consists of a feature embedding layer, multiscale CNN layers for local context extraction, stacked bidirectional RNN layers for global context extraction, fully connected and softmax layers for final joint secondary structure and solvent accessibility classification. Experimental results on the CB$6133$ dataset, the public CB$513$ benchmark, and the recent CASP$10$ and CASP$11$ datasets demonstrate that our proposed deep network outperforms existing methods and achieves state-of-the-art performance.
\item To our knowledge, it is the first time to apply bidirectional GRU layers to secondary protein structure prediction. An ablation study indicates they form the most important component of our deep neural network.
\end{itemize}

The rest of this paper is organized as follows. In Section \ref{sec.model}, we introduce our proposed end-to-end deep model in detail. We present implementation details, experimental results and an ablation study in Section \ref{sec.experiment}. Section \ref{sec.conclusion} concludes this paper with final remarks.

\section{Network Architecture}\label{sec.model}

\begin{figure*}[t]
\centering%
{\includegraphics[width=\textwidth]{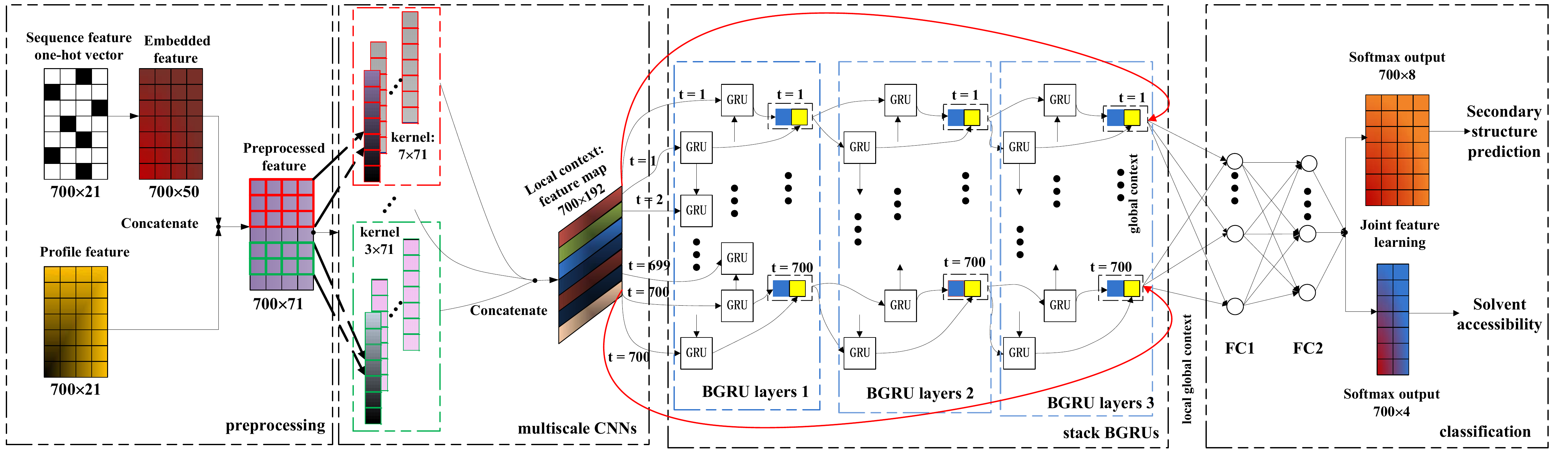}}
\caption{Our end-to-end deep convolutional and recurrent neural network (DCRNN) for predicting protein secondary structures. The input consists of sequence features and profile features. Through feature embedding and concatenation, preprocessed features are fed into the multiscale CNN layer, where multiple kernel sizes are used to extract multiscale local features. The concatenated multiscale features as local contexts flow into three stacked BGRU layers for capturing global contexts. On top of the stacked BGRU layers, two fully connected hidden layers are used for multi-task joint classification.}\label{fig.pipeline}
\end{figure*}

As illustrated in Fig.~\ref{fig.pipeline}, our deep convolutional and recurrent neural network (DCRNN) for protein secondary structure prediction consists of four parts, one feature embedding layer, multiscale convolutional neural network (CNN) layers, three stacked bidirectional gated recurrent unit (BGRU) layers and two fully connected hidden layers. The input to our deep network carries two types of features of a protein amino acid sequence, sequence features and profile features. The feature embedding layer is responsible for transforming sparse sequence feature vectors into denser feature vectors in a new feature space. The embedded sequence features and the original profile features are fed into multiscale CNN layers with different kernel sizes to extract multiscale local contextual features. The concatenated multiscale local contexts flow into three stacked BGRU layers, which capture global contexts. On top of the cascaded CNN and BGRU layers, there are two fully connected hidden layers taking concatenated local and global contexts as input. The output from the second fully connected layer with softmax activation is fed into the output layer, which performs $8-$category secondary structure and $4-$category solvent accessibility classification.

\subsection{Feature Embedding}
For better understanding, protein secondary structure prediction can be formulated as follows. Given an amino acid sequence $X = x_1, x_2, \ldots, x_T$, we need to predict the secondary structure label of every amino acid, $S = s_1, s_2, \ldots, s_T$, where $x_i (\in \mathbb{R}^n)$ is an $n$-dimensional feature vector corresponding to the $i$-th amino acid, and $s_i$ is an 8-state secondary structure label. In this paper, the input feature sequence $X$ is decomposed into two parts, one is a sequence of $21$-dimensional feature vectors encoding the types of the amino acids in the protein and the other is a sequence of $21$-dimensional profile features obtained from the PSI-BLAST \cite{altschul1997gapped} log file and rescaled by a logistic function~\cite{Jones1999}. Note that each feature vector in the first sequence is a sparse one-hot vector, i.e., only one of its 21 elements is none-zero, while a profile feature vector has a dense representation. In order to avoid the inconsistency of feature representations, we adopt an embedding operation from natural language processing to transform sparse sequence features to a denser representation~\cite{mesnil2015using}. This embedding operation is implemented as a feedforward neural network layer with an embedding matrix, $W_{emb} \in \mathbb{R}^{21 \times D_{emb}}$, that maps a sparse $21$-dimensional vector into a denser $D_{emb}$-dimensional vector. In this paper, we empirically set $D_{emb}=50$,
and initialize the embedding matrix with random numbers. The embedded sequence feature vector is concatenated with the profile feature vector before being fed into multiscale CNN layers.

\subsection{Multiscale CNNs}
As illustrated in Fig. \ref{fig.pipeline}, the second component of our deep network is a set of multiscale convolutional neural network layers.
Given the amino acid sequence with embedded and concatenated features
\begin{equation}
\tilde{X} = \left[\tilde{x}_1, \tilde{x}_2, \ldots , \tilde{x}_T \right],
\end{equation}
where $\tilde{x}_i$ ($\in \mathbb{R}^m$) is the preprocessed feature vector of the $i$-th amino acid. To model local dependencies of adjacent amino acids, we leverage CNNs with a sliding window and Rectified Linear Unit (ReLU)~\cite{nair2010rectified} to extract local contexts.
\begin{equation*}
\bar{l}_i = F \ast \tilde{x}_{i:i+f-1} = \text{ReLU}(w \cdot \tilde{x}_{i:i+f-1} + b),
\end{equation*}
where $F$ ($\in \mathbb{R}^{f \times m}$) is a convolutional kernel, $f$ is the extent of the kernel along the protein sequence and $m$ is the feature dimensionality at individual amino acids ($m=71$ in this paper), $b$ is the bias term and `ReLU' is the activation function. The kernel goes through the full input sequence and generates a corresponding output sequence,
$\tilde{L} = \left[\tilde{l}_1, \tilde{l}_2, \ldots, \tilde{l}_T \right]$, where each $\tilde{l}_i$ ($\in \mathbb{R}^q$) has $q$ channels ($q=64$ in this paper).
Since an amino acid is sometimes affected by other residues at a relative large distance, e.g., two residues have interaction at a distance of $11$ in the disulphide bond labeled as link $2$ in Fig. \ref{fig1.secondary}, multiscale CNN layers with different kernel sizes are used to obtain multiple local contextual feature maps. In this paper, we use three CNN layers with $f=3$, $7$, and $11$. This results in three feature maps $\tilde{L}_1, \tilde{L}_2, \tilde{L}_3$. These multiscale features are concatenated together as local contexts $L=\text{concatenate}\{\tilde{L}_1, \tilde{L}_2, \tilde{L}_3\}$.

\subsection{BGRUs}
In addition to local dependencies, long-range dependencies, such as line $1$ in Fig. \ref{fig1.secondary}, also widely exist in amino acid sequences. Multiscale CNNs can only capture dependencies among amino acids separated by a distance not larger than its maximum kernel size.   To capture dependencies across a larger distance, we exploit bidirectional gate recurrent units (BGRUs).

\begin{figure}[t]
\centering%
{\includegraphics[width=0.7\columnwidth]{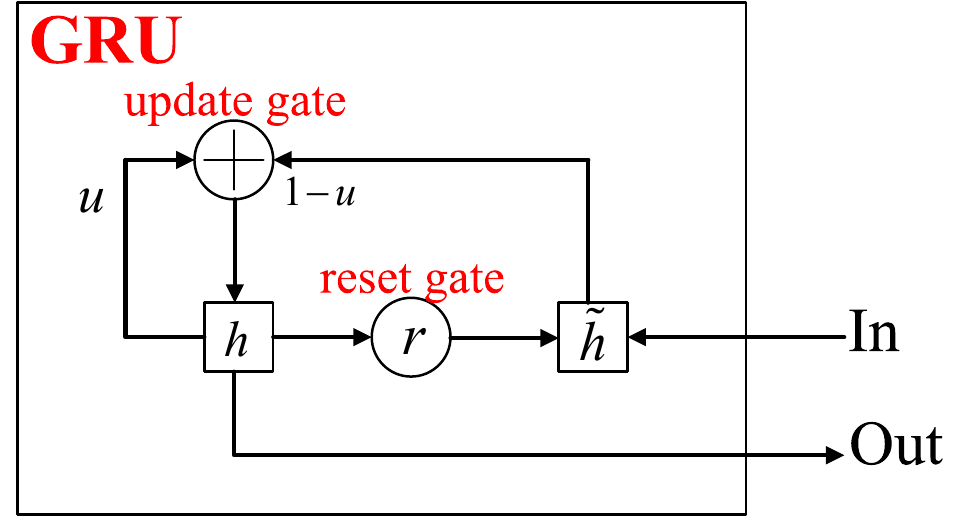}}
\caption{The internal structure of a gate recurrent unit (GRU).}\label{fig.gru}
\end{figure}

Recurrent neural networks (RNNs) have a powerful capacity to deal with context-dependent sequences. However, training RNNs used to be difficult due to vanishing gradients~\cite{Bengio1994}. Only in recent years RNNs with gated units, e.g., long short term memory (LSTM) and gate recurrent units (GRUs), became practically useful. In our network, GRUs~\cite{cho2014properties} are used to capture global contexts because they achieve comparable performance with less parameters in comparison to LSTM~\cite{ilya2015}. In the same notation as previous equations, the mechanism of a GRU (illustrated in Fig. \ref{fig.gru}) can be presented as follows if the input is ($l_t, h_{t-1}$).
\begin{align}
r_t &= \text{sigm} \left( W_{lr} \cdot l_t + W_{hr} \cdot h_{t - 1} + b_r \right),\\
u_t &= \text{sigm}\left( W_{lu} \cdot l_t + W_{hu} \cdot h_{t - 1} + b_u \right),\\
\tilde h_t &= \tanh \left( W_{l\tilde h} \cdot l_t + W_{h\tilde h} \cdot (r_t \odot h_{t - 1} + b_{\tilde h} \right),\\
h_t &= u_t \odot h_{t - 1} + \left(1 - u_t \right) \odot \tilde h_t,
\end{align}
where $r_t, u_t, {{\tilde h}_t}, h_t$ ($\in \mathbb{R}^k$) are respectively activations of the reset gate, update gate, internal memory cell, and GRU output if the number of hidden units is $k$; $W_{lr}, W_{hr}, W_{lu}, W_{hu}, W_{l\tilde h}, W_{h\tilde h}$($\in \mathbb{R}^{3q\times k}$) are weight matrices; and $b_r, b_u, b_{\tilde h}$ ($\in \mathbb{R}^k$) are bias terms. In addition, $\odot$, sigm and $\tanh$ stand for element-wise multiplication, sigmoid and hyperbolic functions, respectively. Compared with LSTM, which has three gates (i.e., input gate, forget gate, output gate), one external memory cell state and one output state, a GRU only has two gates (update gate, reset gate) and one output state. It does not have the least important gate (output gate) in LSTM, and merges the input gate and the forget gate together to form the update gate and the reset gate, which control when information should be artificially remembered or forgotten.
The total number of parameters in a GRU is only $\sfrac{3}{4}$ of that in LSTM~\cite{ilya2015}.

The secondary structure label of an amino acid not only depends on the label of its preceding amino acid in the sequence, but also the label of its following amino acid. Thus, we use bidirectional GRUs, each of which consists of a forward GRU (from $t=0$ to $t=T$)
and a backward GRU (from $t=T$ to $t=0$).
The output from both forward and backward GRUs at time $t$ are concatenated together to form the output from the bidirectional GRU (BGRU) at the same time. Furthermore, to enhance global information flow in our network, three BGRUs are stacked together with dropout to improve the performance. Note that the forward hidden state of the stacked BGRUs is calculated as $h^{z}_{forward\underline~t} =  \text{GRU}(h^{z-1}_t, h^{z}_{forward\underline ~t-1})$, where $z$ and $h^{z-1}_t$ stand for the layer index and the concatenated output of the preceding layer in the stacked BGRUs. At last, the obtained local and global contexts are concatenated together as the input to the following layers.

\subsection{Multi-Task Joint Feature Learning}

Taking the interaction between different protein structure properties into consideration, we train our proposed model to generate two different but correlated types of results by performing joint feature learning in two shared fully connected hidden layers, as shown in the classification part of Fig. \ref{fig.pipeline}. Specifically, the output from our proposed model consists of the predicted sequences of secondary structure labels $s_i$ and solvent accessibility labels $a_i$ (four-category classification problem, i.e., absolute and relative solvent accessibility). Absolute and relative accessibility are judged by specific thresholds of the original and normalized solvent accessibility values computed by the DSSP program \cite{kabsch1983dictionary}, and solvent accessibility is closely related to secondary structure prediction. According to the multi-task training method from \cite{ren2015faster}, $l_2-$norms are added to the loss function as the regularization term. In addition, dropout \cite{srivastava2014dropout} is employed in the stacked BGRU layers and the penultimate layer to avoid overfitting. The joint loss function can be formulated as follows.
\begin{align}\label{loss}
L(\{s_i\},\{a_i\}) = &\frac{1}{N} \sum_i L_s(s_i, s^*_i) \nonumber\\
                     & + \lambda_1 \frac{1}{N} \sum_i L_a(a_i, a^*_i) + \lambda_2 \|\theta\|_2,
\end{align}
where $L_s(s_i, s^*_i) = -s^*_i \log({s_i})$ and $L_a(a_i, a^*_i) = -a^*_i \log({a_i})$ are respective loss functions for secondary structure prediction and solvent accessibility prediction, $s_i$ and $a_i$ are predicted probabilities of secondary structure labels and solvent accessibility labels respectively, $s^*_i$ and $a^*_i$ are ground-truth labels of secondary structure and solvent accessibility respectively, $\theta$ is the weight vector, and $N$ is the number of residues.


\section{Experimental Results}\label{sec.experiment}
\subsection{Datasets and Features}
We use four publicly available datasets, CB$6133$ produced with PISCES CullPDB~\cite{wang2003pisces}, CB$513$~\cite{cuff1999evaluation}~\footnote{\url{http://www.princeton.edu/~jzthree/datasets/ICML2014/}}, CASP$10$~\cite{kryshtafovych2014assessment} and CASP$11$~\cite{moult2014critical}, to evaluate the performance of our proposed deep neural network. CB$6133$ is a large non-homologous protein sequence and structure dataset, that has $6128$ proteins, which include $5600$ proteins (index $0$ to $5599$) for training, $256$ proteins (index $5877$ to $6132$) for validation and $272$ proteins (index $5605$ to $5876$) for testing. Note that the testing set of CB$6133$ is different from that in~\cite{wang2016protein}. CB$513$ is a public benchmark dataset used for testing only. Since there exists redundancy between CB$513$ and CB$6133$,
a smaller filtered version of CB$6133$ is formed by removing sequences in CB$6133$ that have over $25\%$ similarity with some sequence in CB$513$, as done in~\cite{zhoujian2014,wang2016protein}. The filtered CB$6133$ dataset has $5534$ proteins, which can all be used for training if CB$513$ is used as the testing set. CASP$10$ and CASP$11$ contain $123$ and $105$ domain sequences, respectively. They are used for testing the performance of our proposed network trained on the filtered CB$6133$ dataset. The performance of secondary structure prediction is measured by Q$8$ accuracy.

Every protein sequence in the aforementioned datasets has $55$ channels of data per residue. Among the $55$ channels, $21$ channels are used for the sequence feature, which specifies the category of the amino acid, $21$ channels for the sequence profile (PSSM rescaled through a logistic function, PSSM calculated by PSI-BLAST against the UniRef$90$ database with E-value threshold $0.001$ and $3$ iterations), $8$ channels for secondary structure category labels, $2$ channels for solvent accessibility labels (obtained by the DSSP program through $3$D PDB). Although additional features could be considered to further improve performance, we focus on network architecture in this paper. Note that there exist $3$ mask channels after the sequence feature, the sequence profile and the secondary structure label.
For the convenience of subsequent processing and implementation, the length of all protein amino acid sequences in these datasets are normalized to $700$. Sequences longer than $700$ are truncated and those shorter than $700$ are padded with zeros. Most sequences are shorter than $700$.

\subsection{Implementation Details}

In our experiments, multiscale CNN layers with kernel size $3$, $7$, and $11$ are used to extract local contexts given the window size of ``a biological word'' in~\cite{asgari2015protvec} as a reference. The obtained 3 feature maps, each having $64$ channels, are concatenated together as the local contextual feature vector. Each of the three stacked BGRU layers has $600$ hidden units. They take the local contexts as the input. The output from the BGRU layers is regularized with dropout ($=0.5$) to avoid overfitting. The local and global contexts obtained from multiscale CNN layers and BGRU layers are concatenated together and fed to two fully connected layers with ReLU activation. We set $\lambda_1 = 1, \lambda_2=0.001$ for balancing the two joint learning tasks and the regularization term.

We also exploit bagging to obtain ensemble models. According to the standard bagging algorithm, for each weak model, we randomly choose $512$ (about $10\%$) proteins from the original training set to form the validation set and the remaining training samples form the training set. Our ensemble model consists of 10 independently trained weak models. Early stopping is used during training. Specifically, when the F$1$ score on the validation set is not increasing for 10 epoches, we decrease the learning rate by a factor of $2$. Once the learning rate is smaller than a predefined threshold, we stop training, test the model obtained after every epoch on the validation set, and choose the one with the best performance on the validation set as our trained model.

Our code is implemented in Theano~\cite{BastienTheano2012,bergstraal2010scipy}, a publicly available deep learning software\footnote{\url{http://deeplearning.net/software/theano/}}, on the basis of the Keras~\cite{chollet2015} library\footnote{\url{http://keras.io/}}. Weights in our neural networks are initialized using the default setting in Keras. We train all the layers in our deep network simultaneously using the Adam optimizer \cite{kingma2014adam}. The batch size is set to 128. The entire deep network is trained on a single NVIDIA GeForce GTX TITAN X GPU with 12GB memory. It takes about one day to train our deep network without early stopping while it only takes $6$ hours if we take advantage of early stopping. In the testing stage, one protein takes $5$ms on average.

\subsection{Performance}
We evaluate the overall performance of our deep network (DCRNN) by performing three sets of experiments. In the first set of experiments, we perform both training and testing on the original CB$6133$ dataset. In the second set, we perform training on the filtered CB$6133$ dataset and testing on the CB$513$ benchmark. In the third set, we still perform training on the filtered CB$6133$ dataset but performance is measured on the recent CASP$10$ and CASP$11$ datasets.

\begin{table}[t]
\centering
\caption{Classification precision and recall of individual secondary structure labels on the testing set of CB$6133$. Our results (DCRNN) are compared with GSN. Frequencies of secondary structure labels in the training set are given in the last column. Boldface numbers indicate best performance.}\label{table.cb6133}
\begin{tabular}{l|ll|ll|l}
\hline
   & \multicolumn{2}{c|}{Precision}  & \multicolumn{2}{c|}{Recall}     & Frequency \\ \hline
   & DCRNN           & GSN            & DCRNN           & GSN            &           \\ \hline
Q8 & \textbf{0.732} & 0.721          & \textbf{}      &                &           \\
H  & \textbf{0.878} & 0.828          & 0.927          & \textbf{0.935} & 0.344     \\
E  & \textbf{0.792} & 0.748          & \textbf{0.862} & 0.823          & 0.217     \\
L  & \textbf{0.589} & 0.541          & \textbf{0.662} & 0.633          & 0.193     \\
T  & \textbf{0.577} & 0.548          & \textbf{0.572} & 0.506          & 0.113     \\
S  & \textbf{0.518} & 0.423          & \textbf{0.275} & 0.159          & 0.083     \\
G  & 0.434          & \textbf{0.496} & \textbf{0.311} & 0.133          & 0.039     \\
B  & \textbf{0.596} & 0.500          & \textbf{0.049} & 0.001          & 0.010     \\
I  & 0.000          & 0.000          & 0.000          & 0.000          & 0.000
\end{tabular}
\end{table}

\subsubsection{Training and Testing on CB${\bf{6133}}$}
Our model trained using the training set of CB$6133$ achieves ${\bf{73.2}}\pm0.6\%$ Q$8$ accuracy, which defines a new state of the art and is ${\bf {1.1\%}}$ higher than the previous best result obtained with GSN \cite{zhoujian2014}, and $76.1\%$ solvent accessibility accuracy on the testing set of CB$6133$. We did not compare against the result from \cite{wang2016protein} because the testing set used in \cite{wang2016protein} is different from the testing set defined in CB$6133$, and is not publicly available either.
In Table \ref{table.cb6133}, we compare our overall Q$8$ performance as well as the performance on individual secondary structure labels  with the previous best results achieved by GSN~\cite{zhoujian2014}. Apparently, our proposed model achieves higher precision and recall on almost all individual labels. We believe that the better performance is not only due to the power of the neural networks we use, but also owing to the power of the integrated local and global contexts. Specifically, on the four labels (H, E, L and T) with high frequency, our model achieves better performance due to the fact that our model with millions of parameters has higher representation capacity. Nevertheless, our model also performs better than previous models on low frequency labels most likely because of the integrated local and global contexts, which form the core contribution of this paper.

\begin{table*}[t]
\centering
\caption{A comparison of classification precisions of individual secondary structure labels on CB$513$.
Note that CNF was trained with $78$-dimensional features while GSN, DeepCNF and our model (DCRNN) were trained with the standard $42$-dimensional features. The precision of a single model is calculated using the confusion matrix. The precision of an ensemble model is obtained from the voted confusion matrix.
The frequencies of individual labels in the training set and their descriptions are presented in the last two columns. Boldface numbers indicate best performance.} \label{table.cb513}
\begin{tabular}{cccccc|ccc|c|c}
\hline
                        & \multicolumn{5}{c|}{Single Model}                 & \multicolumn{3}{c|}{Ensemble Model} & Frequency & Description  \\ \hline
\multicolumn{1}{l|}{}   & DCRNN           & CNF            & SSpro8 & GSN           & DeepCNF    & DCRNN             & CNF              & SSpro8    &           & \\ \hline
\multicolumn{1}{l|}{Q8} & \textbf{0.694} & 0.633          & 0.511  & 0.664         & 0.683      & \textbf{0.697}   & 0.649            & 0.510     &           & all types \\
\multicolumn{1}{l|}{H}  & 0.836          & \textbf{0.887} & 0.752  & 0.831         & 0.849      & 0.832            & \textbf{0.875}   & 0.752     & 0.309     &  $\alpha-$helix   \\
\multicolumn{1}{l|}{E}  & 0.739          & \textbf{0.776} & 0.597  & 0.717         & 0.748      & 0.753            & \textbf{0.756}   & 0.598     & 0.213     &  $\beta-$strand   \\
\multicolumn{1}{l|}{L}  & 0.573          & \textbf{0.608} & 0.499  & 0.518         & 0.571      & 0.573            & \textbf{0.601}   & 0.500     & 0.211     &  irregular          \\
\multicolumn{1}{l|}{T}  & \textbf{0.549} & 0.418          & 0.327  & 0.496         & 0.530      & \textbf{0.559}   & 0.487            & 0.330     & 0.118     &   $\beta-$turn          \\
\multicolumn{1}{l|}{S}  & \textbf{0.521} & 0.117          & 0.049  & 0.444         & 0.487      & \textbf{0.518}   & 0.202            & 0.051     & 0.098     &   bend           \\
\multicolumn{1}{l|}{G}  & 0.432          & 0.049          & 0.007  & 0.450         & \textbf{0.490}& \textbf{0.429} & 0.207            & 0.006     & 0.037     & $3_{10}-$helix            \\
\multicolumn{1}{l|}{B}  & \textbf{0.558} & 0.000          & 0.000  & 0.000         & 0.433      & \textbf{0.554}   & 0.000            & 0.000     & 0.014     &  $\beta-$bridge           \\
\multicolumn{1}{l|}{I}  & 0.00           & 0.000          & 0.000  & 0.000         & 0.000      & 0.000            & 0.000            & 0.000     & 0.000     &  $\pi-$helix
\end{tabular}
\end{table*}

\subsubsection{Training on Filtered CB${\bf{6133}}$ and Testing on CB${\bf{513}}$}
We have also performed validation on the public CB$513$ benchmark using a model trained on the filtered CB$6113$ dataset, whose $5534$ proteins do not include any sequences with more than $25\%$ similarity with proteins in CB$513$. Our single trained model achieves ${\bf{69.4}} \pm 0.5 \%$ Q$8$ accuracy, which is ${\bf{1.1}}\%$ higher than the previous state of the art achieved by DeepCNF~\cite{wang2016protein} on the same training and testing sets, and $76.8\%$ solvent accessibility accuracy on the validation set as ground-truth solvent accessibility labels are unavailable on CB$513$. We also compare our model against other existing methods (e.g., CNF~\cite{wang2011}, SSpro$8$~\cite{pollastri2002} and GSN~\cite{zhoujian2014}) on individual secondary structure labels in Table \ref{table.cb513}.
Note that, in addition to the standard sequence feature and profile feature used by other methods (including ours) in this comparison, the CNF model~\cite{wang2011} was trained using three extra features, and the entire feature vector at an amino acid is $78$-dimensional. Even though CNF was trained using more features, our model still achieves an overall higher Q$8$ accuracy. In terms of individual labels, our model achieves slightly lower accuracies ($3\%$ to $4\%$ difference) on high frequency labels (H, E and L) and significantly higher accuracies on low frequency labels (T, S and B).

In order to further improve accuracy and robustness, we also compute an ensemble model by averaging $10$ weak models trained independently on 10 randomly sampled training and validation subsets according to the bagging algorithm. Through the model averaging process, the Q$8$ accuracy can be improved to ${\bf{69.7\%}}$ (${\bf{4.8}}\%$ higher than the previous best result from ensemble models), as shown in Table~\ref{table.cb513}. In addition, the Q$3$ accuracy of our single model is ${\bf{84.0\%}}$, which is ${\bf{1.7\%}}$ higher than the previous state of the art ($82.3\%$)~\cite{wang2016protein}. The mapping between $8$-state labels and $3$-state labels is as follows: H ($8$-state) is mapped to H ($3$-state), E ($8$-state) is mapped to E ($3$-state) and all other $8$-state labels are mapped to C ($3$-state) according to \cite{wang2016protein}. In addition, the p-value of the significance test, ``our model outperforms other methods'', is $1.4\times10^{-5}$ ($< 0.001$) using results from 10 different runs.

\subsubsection{Training on Filtered CB${\bf{6133}}$ and Testing on CASP${\bf{10}}$ and CASP${\bf{11}}$}
For further verifying the generalization capability of our model trained on the filtered CB$6133$ dataset, we have also evaluated it on more recent datasets, CASP$10$ and CASP$11$. We compare the Q$8$ accuracy of our model against SSpro~\cite{magnan2014sspro}, RaptorX-SS$8$~\cite{wang2011} and DeepCNF. In addition, the Q$3$ accuracy of our model is also compared against SSpro, SPINE-X~\cite{faraggi2012spine}, PSIPRED~\cite{Jones1999}, RaptorX-SS$8$, JPRED~\cite{drozdetskiy2015jpred4} and DeepCNF. According to the results shown in Table~\ref{table.casp}, the Q$8$ accuracy of our model is ${\bf{76.9\%}}$ over CASP$10$ (${\bf{5.1\%}}$ higher than pervious best) and ${\bf{73.1\%}}$ over CASP$11$ (${\bf{0.8\%}}$ higher than pervious best). The Q$3$ accuracy of our model over CASP$10$ and CASP$11$ is respectively ${\bf{87.8\%}}$ (${\bf{3.4\%}}$ higher than previous best) and ${\bf{85.3\%}}$ (${\bf{0.6\%}}$ higher than previous best). Note that the Q$8$ accuracy and Q$3$ accuracy of our method in this test are obtained through a single model rather than an ensemble model for fair comparison.
\begin{table}[t]
\centering
\caption{Training on filtered CB${\bf{6133}}$ and testing on CASP${\bf{10}}$ and CASP${\bf{11}}$. Both Q$8$ and Q$3$ accuracies are reported for SSpro, SPINE-X, PSIPRED, JPRED, Raptorx-SS$8$, DeepCNF and our model (DCRNN). Boldface numbers indicate best performance.}
\label{table.casp}
\resizebox{0.45\textwidth}{!}
{
\begin{tabular}{l|c|c|c|c}
\hline
                                     & \multicolumn{2}{c|}{\textbf{Q8(\%)}}                      & \multicolumn{2}{c}{\textbf{Q3(\%)}}                     \\ \cline{2-5}
\multicolumn{1}{c|}{\textbf{Method}} & \multicolumn{1}{l|}{CASP10} & \multicolumn{1}{l|}{CASP11} & \multicolumn{1}{l|}{CASP10} & \multicolumn{1}{l}{CASP11} \\ \hline
SSpro(without template)              & 64.9                        & 65.6                        & 78.5                        & 77.6                       \\ \hline
SSpro(with template)                 & 75.9                        & 66.7                        & 84.2                        & 78.4                       \\ \hline
SPINE-X                              & -                           & -                           & 80.7                        & 79.3                       \\ \hline
PSIPRED                              & -                           & -                           & 81.2                        & 80.7                       \\ \hline
JPRED                                & -                           & -                           & 81.6                        & 80.4                       \\ \hline
Raptorx-SS$8$                        & 64.8                        & 65.1                        & 78.9                        & 79.1                       \\ \hline
DeepCNF                              & 71.8                        & 72.3                        & 84.4                        & 84.7                       \\ \hline
DCRNN                                & \textbf{76.9}               & \textbf{73.1}               & \textbf{87.8}               & \textbf{85.3}              \\ \hline
\end{tabular}
}
\end{table}

\subsection{Ablation Study}
To discover the vital elements in the success of our proposed network, we conduct an ablation study by removing or replacing individual components in our network. Specifically, we have tested the performance of models without the feature embedding layer, multiscale CNNs, stacked BGRUs or backward RNN passes. In addition, we also tested a model that does not feed local contexts to the last two fully connected layers, and another one where the BGRU layers are replaced with bidirectional simpleRNN layers without any gate structure to figure out the importance of the gate structure in presence of long-range dependencies. From the results on the CB$513$ dataset presented in Table \ref{table.comparative}, we find that those bidirectional GRU layers are the most effective component in our network as the performance drops to $66.9\%$ when we run forward RNN passes only without backward RNN passes.  Multiscale CNNs are also important as the performance drops to $68.1\%$ without them. In addition, compared with bidirectional simpleRNN, the gate structure in our GRU layers is necessary for dealing with long-range dependencies widely existing in amino acid sequences. Stacked BGRU layers are also beneficial for enhancing global information circulation when compared with a single BGRU layer. Furthermore, directly feeding local contexts to the fully connected layers in addition to the global contexts is also essential for good performance, especially for the prediction of low-frequency secondary structure categories. Last but not the least, feature embedding, multi-task learning and bagging can all be applied to improve the accuracy and robustness of our method.
\begin{table}[t]
\renewcommand{\arraystretch}{1}
\caption{An ablation study on CB$513$.}
\centering
\begin{tabular}{l c}
  \hline
  \multicolumn{1}{c}{Model}  & Q$8$ Accuracy \\
  Without feature embedding& $68.9\%$ \\
  Without multiscale CNNs & $68.1\%$ \\
  Without backward GRU passes& $66.9\%$ \\
  Replacing GRU with simpleRNN & $68.0\%$ \\
  Single BGRU layer & $68.6\%$ \\
  Without integrated local\&global contexts & $68.8\%$\\
  Without Multi-task learning & $68.8\%$ \\
  \hline
\end{tabular}\label{table.comparative}
\end{table}

\section{Conclusions}\label{sec.conclusion}
To apply recent deep neural networks to protein secondary structure prediction, we have proposed an end-to-end model with multiscale CNNs and stacked bidirectional GRUs for extracting local and global contexts. Through the integrated local and global contexts, the previous state of the art in protein secondary structure prediction has been improved. By taking interactions between different protein properties into consideration, multi-task joint feature learning is exploited to further refine the performance. Because of the success of our proposed deep neural network on secondary structure prediction, such models combining local and global contexts can be potentially applied to other challenging structure prediction tasks in protein and computational biology.


Existing BGRUs are unable to deal with extremely long dependencies, especially low frequency long dependencies. More powerful architectures with an implicit attention mechanism, such as neural Turing machines~\cite{Grave2014}, may be suitable for solving this problem and further improving the prediction performance.

\section*{Acknowledgments}
We wish to thank Sheng Wang and Jian Zhou for discussion and consultation about datasets, and the anonymous reviewers for their valuable comments. The first author is supported by Lee Shau Kee Postgraduate Fellowship from the University of Hong Kong.

\bibliographystyle{named}
\bibliography{ijcai16_lz}

\end{document}